\documentclass[trackchanges]{aastex701}

\usepackage{amsmath}
\usepackage{ulem} 
\usepackage{xcolor}

\begin{document}

\title{Evidence for candidate X-ray pulsations from the ultraluminous X-ray source NGC 7456 ULX-1}

\author[0009-0004-7882-7777]{Yuanle Yao}
\affiliation{School of Astronomy and Space Science, Nanjing University, Nanjing 210023, P. R. China}
\affiliation{Key Laboratory of Modern Astronomy and Astrophysics (Nanjing University),
Ministry of Education, Nanjing 210023, P. R. China}
\email{yaoyl@smail.nju.edu.cn}

\author[orcid=0000-0002-0584-8145]{Xiang-Dong Li}
\affiliation{School of Astronomy and Space Science, Nanjing University, Nanjing 210023, P. R. China}
\affiliation{Key Laboratory of Modern Astronomy and Astrophysics (Nanjing University),
Ministry of Education, Nanjing 210023, P. R. China}
\email[show]{lixd@nju.edu.cn}

\author[orcid=0000-0002-3614-1070]{Xiao-Jie Xu}
\affiliation{School of Astronomy and Space Science, Nanjing University, Nanjing 210023, P. R. China}
\affiliation{Key Laboratory of Modern Astronomy and Astrophysics (Nanjing University),
Ministry of Education, Nanjing 210023, P. R. China}
\email{xuxj@nju.edu.cn}

\begin{abstract}
We report evidence for a candidate pulsational signal at $\sim0.22$~Hz from NGC~7456~ULX-1, a previously identified ultraluminous X-ray source (ULX). The signal is identified in the 2023 \textit{XMM-Newton} observation using independent timing techniques including accelerated searches, $Z^2_n$ statistics, and an orbital-demodulation analysis designed to restore phase coherence in the presence of binary motion. The candidate pulsation frequency drift within the observation suggests rapid spin evolution driven by accretion torque. We further estimate the surface dipole magnetic field strength to be $B\sim 10^{12}$--$10^{14}$~G. These results provide evidence that NGC~7456~ULX-1 may host an accreting neutron star, although confirmation with independent datasets or additional observations is required.
\end{abstract}


\section{\textbf{Introduction}} \label{sec:intro}

Ultraluminous X-ray sources (ULXs) are point-like, off-nuclear, and predominantly extragalactic X-ray sources with luminosities exceeding $10^{39}$ erg\,s$^{-1}$, well above the Eddington limit for both neutron stars and typical stellar-mass black holes \citep{kaaret_ultraluminous_2017}. The nature of ULXs has been a subject of intense debate, with two leading hypotheses: (1) that these systems harbor intermediate-mass black holes (IMBHs), or (2) that they consist of stellar-mass black holes or neutron stars undergoing super-Eddington accretion \citep[][for recent reviews]{Fabrika2021,king_ultraluminous_2023}.

A subset of ULXs, known as ultraluminous X-ray pulsars (ULXPs), exhibit coherent X-ray pulsations, confirming the presence of neutron stars as central accreting objects. The first detection of pulsations from M82 X-2 marked a pivotal breakthrough in ULX research \citep{bachetti_ultraluminous_2014}, and since then, approximately a dozen ULXPs have been identified in nearby galaxies \citep[see Table 3 in][and references therein]{king_ultraluminous_2023}.
However, the vast majority of the over 1800 ULX candidates discovered to date are located at substantial distances ($\gtrsim 5\, \mathrm{Mpc}$) \citep{walton_multi-mission_2021, bernadich_expanded_2022}.
Their low flux levels ($\lesssim 10^{-13}\, \mathrm{erg\;s^{-1}\,cm^{-2}}$), Doppler modulation induced by orbital motion, and  rapid intrinsic spin evolution due to rapid accretion make pulsation searches far more challenging than for conventional X-ray pulsars.

In recent years, advances in search methodologies have enabled the detection of faint and transient pulsational signals in several ULXs. Notable examples include NGC 2403 XMM4 \citep{luangtip_ngc_2024}, NGC 7793 ULX-4 \citep{quintin_new_2021}, and NGC 1313 X-2 \citep{sathyaprakash_discovery_2019}, where pulsations were detected despite their challenging observational conditions. The confirmation of these pulsations strengthens the hypothesis that a substantial fraction of ULXs may host neutron stars as their central accreting objects, rather than IMBHs.

NGC 7456 is a barred spiral galaxy located in the constellation Grus, approximately 15.7 Mpc from Earth \citep{Tully_2016}. This galaxy hosts several ULXs, including ULX-1, ULX-2, ULX-3, ULX-4 \citep{walton_2xmm_2011}, as well as a recently discovered candidate ULX-5 \citep{pintore_ultraluminous_2020}. Among these, ULX-1 is the most luminous, with a peak luminosity reaching $\sim 10^{40}\,\mathrm{erg\;s^{-1}}$.  Its X-ray spectra can be described by a model with two thermal components, as often found in ULXs \citep{pintore_ultraluminous_2020}. Since no pulsations were detected, its nature remains unclear.

 In this study, we present the results of our pulsation search in ULX-1.
The structure of this paper is organized as follows: Section 2 details the observational setup and data reduction procedures. Section 3 describes our data analysis methods and presents the results. Finally, Section 4 discusses the scientific implications of our findings.

\begin{figure}[ht!]
\plotone{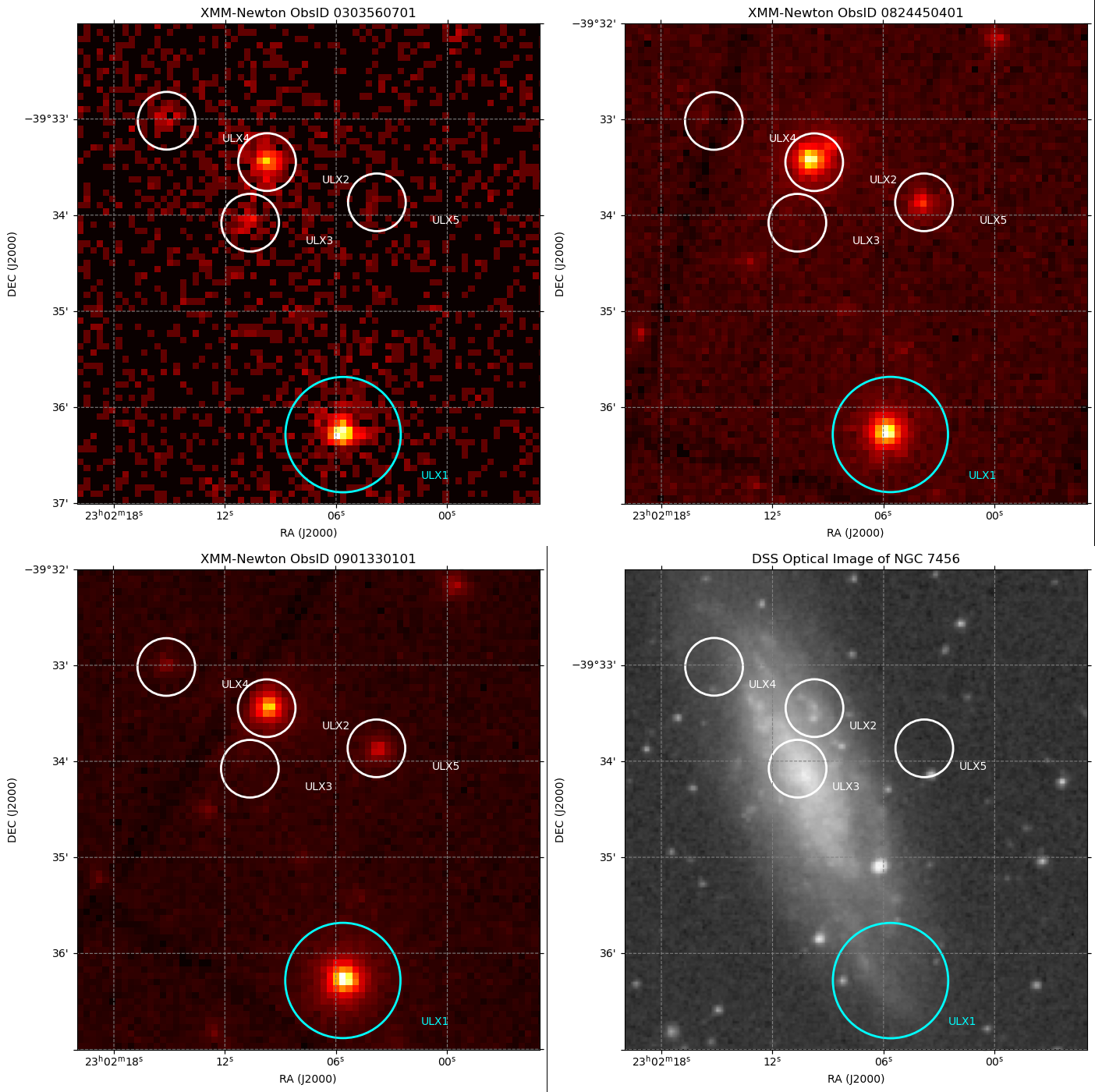}
\caption{Comparison of XMM-Newton X-ray images of NGC 7456 from 2005 (upper left), 2018 (upper right), and 2023 (lower left) with a DSS optical image (lower right). The cyan circle denotes the location of ULX-1, while the white circles indicate the positions of other ULXs.
\label{fig:general}}
\end{figure}

\begin{deluxetable*}{ccccccc}[ht!]
\tabletypesize{\scriptsize}
\tablewidth{0pt}
\tablecaption{XMM-Newton observational details and candidate timing parameters of NGC 7456 ULX-1 \label{tab:xmmObs}}
\tablehead{
Obs.ID. & Date & Total Expos (ks) & $L_{\rm X} (10^{39} {\rm erg\,s}^{-1})$ &
Period (s) & $\dot{P}$ ($10^{-8}$ s\,s$^{-1}$) & $\dot{\nu}$ ($10^{-9}$ Hz\,s$^{-1}$)
}
\startdata
0303560701 & 2005-05-06 & 10.2 & $10.0\pm 1.0$ & -- & -- & -- \\
0824450401 & 2018-05-18 & 92.4 & $5.9\pm 0.4$ &
-- & -- & -- \\
0901330101 & 2023-04-28 & 123.9 & $11.8\pm0.3$ &
$4.51\pm0.03$ & $-8.2\pm0.7$ & $4.0\pm0.3$ \\
\enddata
\tablecomments{
    Luminosity values for the 2005 and 2018 observations from \citep{pintore_ultraluminous_2020},
    with the 2023 luminosity determined via energy spectrum fitting.
    The frequency derivatives are calculated with $\dot{\nu} = -\dot{P}/P^2$.
    The timing parameters listed in this table are representative estimates from the initial timing search and provided for reference only.
}
\end{deluxetable*}

\section{\textbf{Observations and Data Reduction}}

\begin{figure*}[ht!]
\plotone{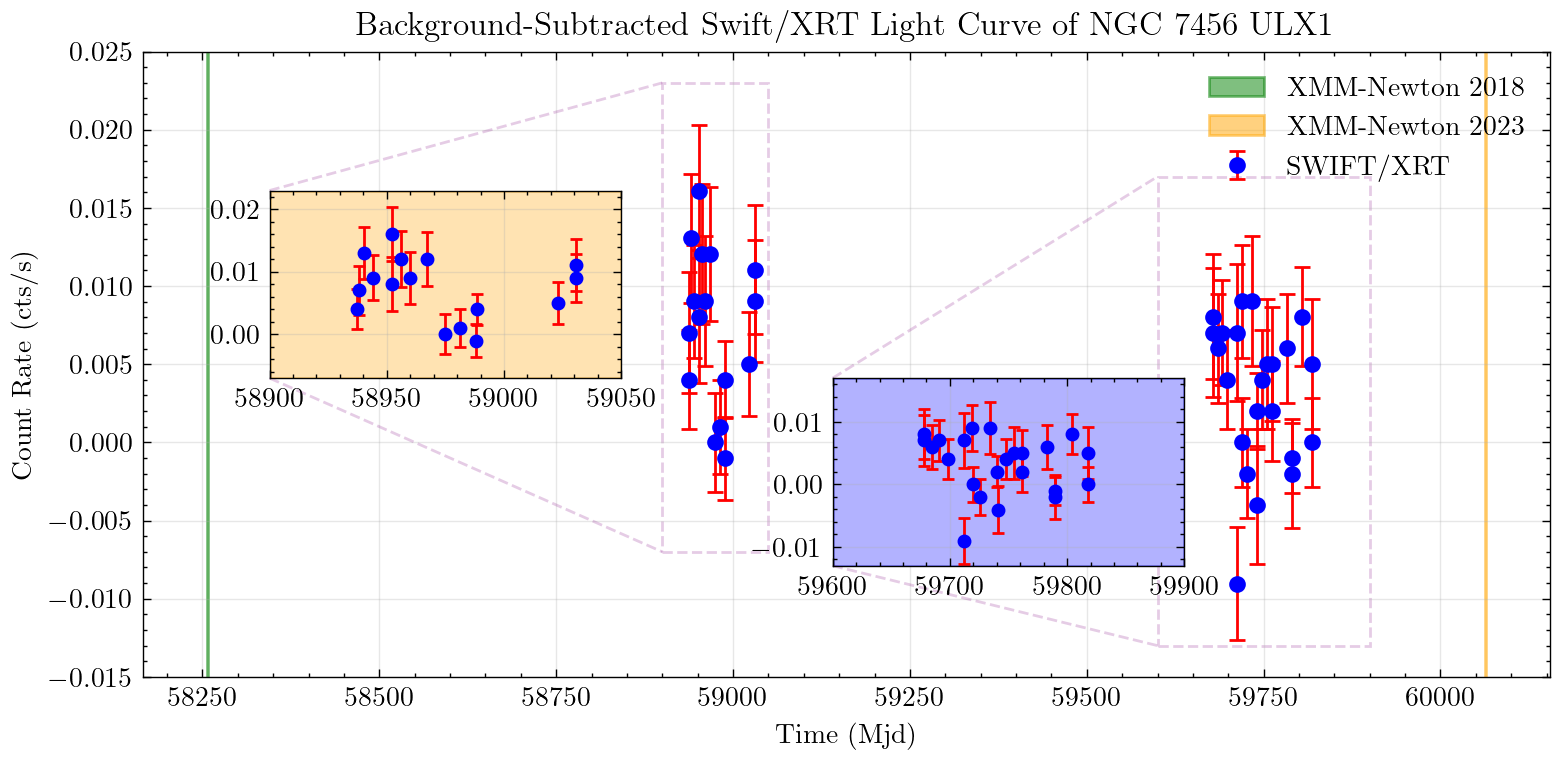}
\caption{Background-subtracted Swift/XRT light Curve of NGC 7456 ULX-1 using a 36-arcsecond extraction region and 800 s time bins.
\label{fig:xrt_lc}}
\end{figure*}

\subsection{XMM-Newton Observations}

The XMM-Newton telescope observed the galaxy NGC 7456 in 2005, 2018, and 2023 \citep{walton_2xmm_2011, pintore_ultraluminous_2020}, with a total nominal exposure of about 226 ks (Table \ref{tab:xmmObs}). The observational data were processed using the XMM-Newton Scientific Analysis System (SAS) v.21.0.0\footnote{\url{https://www.cosmos.esa.int/web/xmm-newton/xsa}}. Data reduction was performed for the EPIC cameras (PN, MOS1, and MOS2), all operating in full-frame imaging mode with a thin filter. Event selection adhered to standard criteria: $\mathrm{FLAG = 0}$, with $\mathrm{PATTERN \le 4}$ for the PN detector, and $\mathrm{PATTERN \le 12}$ for the MOS1 and MOS2 detectors.

For image and spectral analysis, data from both MOS and PN detectors were utilized. However, due to the PN detector’s superior effective area and finer time resolution ($\sim 73$ ms), only the PN data were employed for timing analysis and pulsation searches. Some observations were impacted by strong background flares, which were filtered out using SAS-recommended thresholds: events with count rates exceeding $0.35 \;\mathrm{counts\,s^{-1}}$ (MOS) and $0.4 \;\mathrm{counts\,s^{-1}}$ (PN) were excluded. The resulting cleaned event files were barycenter-corrected to the solar system barycenter using the \texttt{barycen} command.

Figure \ref{fig:general} presents a comparison of XMM-Newton X-ray images (2005, 2018, and 2023) with a DSS optical image\footnote{\url{https://irsa.ipac.caltech.edu/data/DSS/}}. The observational details and representative candidate timing parameters are summarized in Table~\ref{tab:xmmObs}. Photon events were extracted from a circular region with a 36-arcsecond radius centered on the ULX-1 coordinates using the \texttt{evselect} command, and the resulting photon list was used for the pulsation search. {
For reference, the extracted EPIC-pn event lists used in our timing analysis contain
$N_{\gamma}=15113$ (0.2--12.0~keV) and $N_{\gamma}=10826$ (0.5--12.0~keV) events in the 2023 observation,
and $N_{\gamma}=4378$ (0.2--12.0~keV) and $N_{\gamma}=3116$ (0.5--12.0~keV) events in the 2018 observation,
after applying the standard filtering and GTIs described above.
}

\subsection{Swift/XRT Observations}
In addition to XMM-Newton, Swift/XRT telescope observed NGC 7456 in the soft X-ray band in 2020 and 2022, detecting several ULXs, including ULX-1. We used the \texttt{xselect} command to extract photons within a 36-arcsecond radius around the source from Swift/XRT observations of NGC 7456. A light curve with a bin size of 800 seconds was then generated (see Figure \ref{fig:xrt_lc}).

Notably, the luminosity of ULX-1 exhibits significant long-term variability between the 2020 and 2022 observations, which may be related to orbital or superorbital modulation. However, due to the low photon count rate and limited observational coverage, we were unable to constrain a definitive periodicity.

\section{\textbf{Analysis and Results}}

\subsection{Initial Blind Search}

We performed a targeted pulsation search on the 2023 \textit{XMM-Newton} observation of NGC~7456~ULX--1, focusing on the frequency range from 0.01~Hz to the Nyquist frequency (6.8~Hz) using the FTOOL task \texttt{POWSPEC}.
The analysis utilized light curves extracted from predefined energy bands (Section~2) and adopted Leahy normalization \citep{leahy_searches_1983}, which sets the expected Poisson noise level to 2 and provides a convenient framework for identifying periodic signals.

The initial power spectrum did not reveal any prominent narrow peaks.
However, given the potential frequency drift and signal broadening expected from rapid spin evolution and Doppler modulation in ULXs, we proceeded with more sensitive search techniques, including acceleration searches and $Z_n^2$-based methods.

{
To perform an acceleration search, we converted the barycenter-corrected EPIC-pn event list from the 2023 observation into a format compatible with the \texttt{PRESTO} pulsar-search software \citep{ransom_presto_2011} using the \texttt{HENBINARY} command from the \texttt{HENDRICS} package (version~8.1; \citealt{bachetti_extending_2021}). Photon events in the 0.2--12.0~keV band were selected, adopting the intrinsic pn time resolution of 0.074~s. The resulting time series was uniformly binned and Fourier transformed using \texttt{realfft}.
}

{
We then applied the \texttt{ACCELSEARCH} algorithm \citep{ransom2002fourier} to conduct a Fourier-domain acceleration search over the frequency range 0.01--6.8~Hz, summing a single harmonic (\texttt{-numharm 1}) and allowing a maximum Fourier frequency drift of $|z|\leq50$ bins.
This setup corresponds to a maximum line-of-sight acceleration of $\sim5.4~\mathrm{m\,s^{-2}}$ at the detected frequency and involves approximately $1.1\times10^{7}$ independent trial points in the $(\nu,\dot{\nu})$ parameter space.
}

{
The most significant candidate identified in this blind acceleration search occurs at a frequency of $\sim0.22$~Hz.
For this candidate, the detection statistic reported by \texttt{ACCELSEARCH} corresponds to a very small single-trial false-alarm probability ($p_{\rm single}\sim10^{-8}$), indicating a strong local excess at this frequency. After accounting for the large number of independent trials explored in the blind search, the corresponding search-level false-alarm probability is $p_{\rm search}\sim4\times10^{-3}$. While this is insufficient to claim a statistically robust candidate detection on its own, it provides a well-motivated candidate for subsequent targeted and fully
coherent analyses.
}

{
Nevertheless, the presence of a coherent peak in the acceleration-corrected Fourier spectrum motivated further, more targeted timing analyses.
In the following sections, we investigate this candidate using independent techniques, including a refined $Z_n^2$ search and a fully coherent analysis incorporating orbital demodulation.
}

{
To refine the candidate identified in the acceleration search, we performed a targeted $Z_n^2$ analysis using the \texttt{HENDRICS} task \texttt{HENZSEARCH} \citep{bachetti_extending_2021}.
This analysis was carried out over a narrow frequency interval encompassing the candidate region ($0.221$--$0.223$~Hz), and therefore represents a follow-up refinement rather than an independent blind search.
Photon events in the 0.2--12.0~keV band were used, consistent with the timing analysis described above.}

{
We evaluated the $Z^2_n$ statistic for $n_{\rm harm}=1,2,$ and $3$.
In all cases, a clear and isolated peak is recovered near $\nu\simeq0.222$~Hz, with mutually consistent best-fit frequencies and frequency derivatives.
For the 0.2--12.0~keV selection, the maximum values increase monotonically from $Z^2_1\simeq40.6$ to $Z^2_2\simeq44.0$ and $Z^2_3\simeq47.9$, indicating the presence of a predominantly sinusoidal pulse profile with a modest contribution from higher harmonics.
}

{
When restricting the analysis to the 0.3--10.0~keV energy range, the detection strength further increases, with $Z^2_3$ exceeding 50 at the same frequency.
This behavior is consistent with an energy-dependent pulsed fraction and suggests that the candidate pulsation signal is more strongly concentrated at intermediate to hard X-ray energies.
}

\begin{figure}[ht!]
\plotone{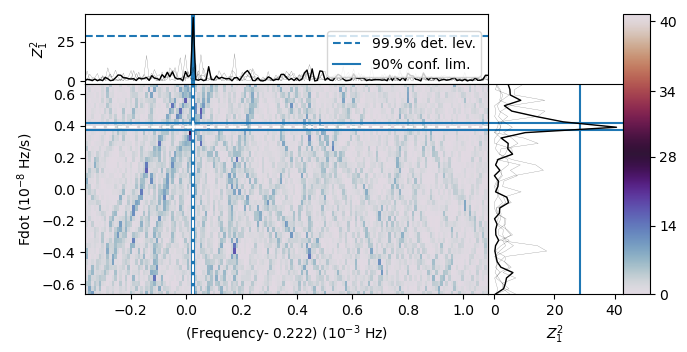}
\caption{
$Z^2_1$ periodogram of the 2023 \textit{XMM-Newton}/PN data obtained using \texttt{HENZSEARCH}.
A clear and isolated peak is detected near $\nu \simeq 0.222$~Hz.
Consistent peaks with higher detection strength are also recovered when additional harmonics are included (see text).
}
\label{fig:zsearch_23}
\end{figure}

{
In the following, we adopt the $Z^2_1$ periodogram to illustrate the refinement results, as it represents the most conservative statistic with the smallest number of degrees of freedom, while higher-harmonic results are used as a consistency check on the robustness of the signal.
The corresponding $Z^2_1$ periodogram for the 2023 observation is shown in Figure~\ref{fig:zsearch_23}.}

{
As an additional consistency check, we applied the same pulsation-search procedures to a nearby ULX (ULX--2) located on the same EPIC-pn detector, as well as to source-free background regions.
Using identical event selections and search configurations, including acceleration searches with \texttt{zmax} values up to 1200, we did not detect any significant periodic signal in either the comparison source or the background.
This result disfavors an instrumental or detector-wide origin for the candidate signal detected in ULX--1. However, ULX--2 contains only about one third as many photons as ULX--1 in the relevant time series, so a weak signal could remain undetected. We therefore regard the ULX--2 search as a consistency check rather than a definitive proof that the ULX--1 candidate is not spurious.
}

Following the identification of the 2023 candidate, we revisited the archival 2005 and 2018 observations.
{
For the 2018 dataset, we conducted an extensive $Z^2_n$ search over a broad parameter space, allowing frequency derivatives corresponding to \texttt{zmax} values up to 1200.
Under these conditions, candidate peaks can be found in the frequency range $0.18$--$0.23$~Hz; however, representative solutions imply extremely large frequency and period derivatives (e.g., $\dot{P}\sim -3.5\times10^{-6}$~s\,s$^{-1}$), which are difficult to reconcile with physically plausible spin evolution in ULX pulsars.
}

The standard flare-filtering procedure for the EPIC-pn detector (excluding intervals with 10--12~keV count rates exceeding $0.4~\mathrm{counts\,s^{-1}}$) resulted in fragmented GTIs in the long 2018 exposure.
By relaxing this threshold to $0.6~\mathrm{counts\,s^{-1}}$, two longer continuous segments (32~ks and 19~ks) were recovered without significantly altering the source light-curve morphology.
{
Even under these optimized conditions, the 2018 data do not yield a unique and physically well-constrained pulsation solution.
We therefore do not present a $Z^2_n$ periodogram for the 2018 observation and do not treat these peaks as secure candidate detections.
}


The representative timing parameters of the 2023 candidate are summarized in Table~\ref{tab:xmmObs}; no secure timing solution is reported for the 2018 observation.

\subsection{Orbital-Corrected Coherent Search}

{
To further test the robustness of the candidate pulsation signal and to investigate whether phase modulation induced by binary motion can account for the reduced significance in the blind and acceleration searches, we performed a fully coherent search on orbit-corrected photon arrival times for the 2023 \textit{XMM-Newton}/PN observation.
}

{
Motivated by the apparent phase wandering of the pulse signal, we adopted a circular-orbit delay model and searched for the set of demodulation parameters that maximizes phase coherence across the full observation. The search was carried out in the five-dimensional parameter space $(a_1, P_{\rm orb}, T_{\rm asc}, \nu, \dot{\nu})$, where $a_1$ is the projected semi-major axis, $P_{\rm orb}$ the orbital period, and $T_{\rm asc}$ the epoch of ascending node passage.
}


The explored parameter ranges were deliberately chosen to be broad:
$a_1 = 1$--$90~\mathrm{s}$,
$P_{\rm orb} = 35$--$500~\mathrm{ks}$,
$T_{\rm asc} = T_0 \pm 500~\mathrm{ks}$, and
$\dot{\nu} = \pm\,2\times10^{-8}~\mathrm{Hz\,s^{-1}}$,
where $T_0$ denotes the first time point of the observation. The adopted $a_1$ range includes very compact projected orbits, since values of only a few seconds are observed in some PULXs \citep[e.g., NGC~5907~ULX--1,][]{Belfiore2024}. To efficiently explore this large parameter space, it was divided into multiple overlapping subregions.
Following previous applications of particle-swarm optimization in pulsation searches \citep{sacchi_restless_2024, pintore_new_2025}, within each subregion, we employed a particle-swarm optimization (PSO) algorithm \citep{pso_1995} as a numerical optimizer. For each trial parameter set, the photon arrival times were demodulated according to the assumed orbital solution, and a coherent periodicity test was performed. Following the orbital-parameter maps presented by \citet{pintore_new_2025}, we then constructed a two-dimensional ``banana-plot" in the $P_{\rm orb}$--$a_1\sin i$ plane by profiling over the remaining phase-correction parameters.

{
As the objective function of the optimization, we directly used the $H$-test statistic \citep{jager_h_test_2010}, allowing up to three harmonics. The $H$-test is defined as
\[
H = \max_{1 \le m \le 3} \left( Z^2_m - 4m + 4 \right),
\]
where $Z^2_m$ is the standard $Z^2$ statistic computed by including the fundamental frequency and its first $(m-1)$ harmonics. In this way, the $H$-test adaptively selects the number of harmonics that maximizes the detection significance without assuming a fixed pulse shape. In the following, $m$ denotes the number of harmonics at which the $H$-test reaches its maximum.
}

{
This procedure consistently converged to a candidate coherent solution near $\nu \simeq 0.2266~\mathrm{Hz}$. The maximum value of the $H$-test reached $H_{\rm max}=64.6$ (maximized at $m=1$). Given the finite grid resolution and the strong covariance among the orbital parameters, we quote only representative values rather than statistically meaningful high-precision digits:
\[
a_1 \simeq 84~\mathrm{s}, \quad
P_{\rm orb} \simeq 1.21\times10^{5}~\mathrm{s}\;(1.40~\mathrm{d}), \quad
\nu \simeq 0.2266~\mathrm{Hz}, \quad
\dot{\nu} \simeq 1.5\times10^{-9}~\mathrm{Hz\,s^{-1}}.
\]
The best-fit orbital period is close to the $\simeq124~\mathrm{ks}$ duration of the 2023 time series. The inferred orbital parameters should therefore be interpreted as an effective phase-correction solution that maximizes coherence over the observation, rather than as a unique physical orbital ephemeris.
}

{
We also evaluated the dependence on photon selection.
For the 2023 dataset, restricting the events to 0.5--12.0~keV yields a noticeably
stronger detection statistic than using 0.2--12.0~keV, indicating that the candidate pulsed
signal is more strongly concentrated at higher energies.
In terms of photon statistics, we used $N_{\gamma}=15113$ (0.2--12.0~keV) and
$N_{\gamma}=10826$ (0.5--12.0~keV) events in 2023.
}

\begin{figure*}[t]
\centering
\includegraphics[width=0.95\textwidth]{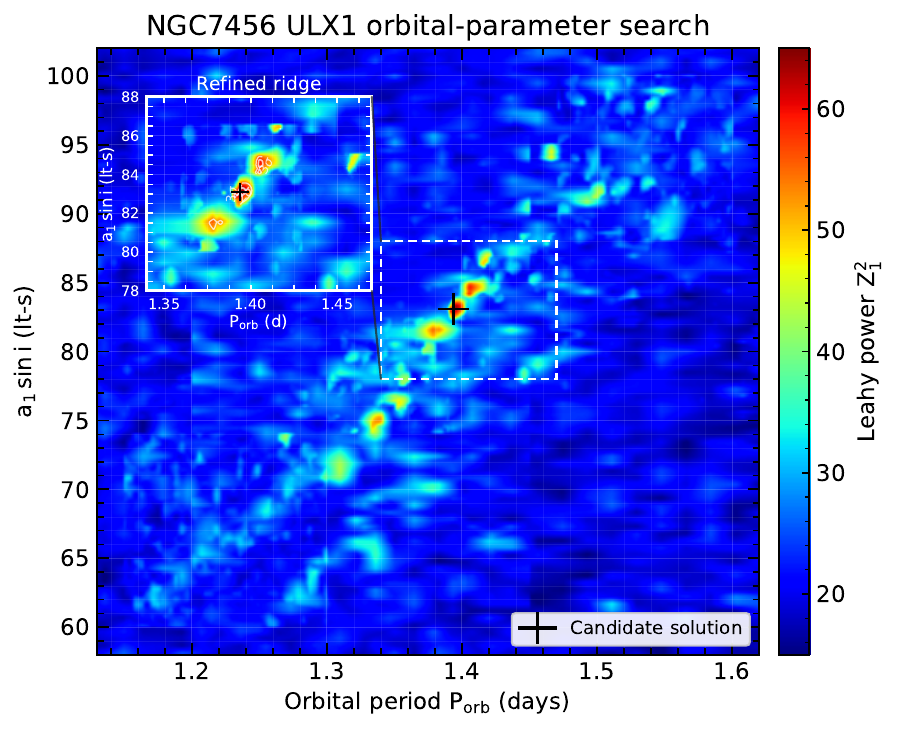}
\caption{Single-harmonic Leahy/Rayleigh orbital-parameter map for the 2023 \textit{XMM-Newton}/PN observation, following the banana-plot representation of \citet{pintore_new_2025}. The main panel shows the extended search in $P_{\rm orb}$ and $a_1\sin i$, and the inset zooms into the refined high-power ridge. At each grid point, we applied a circular-orbit delay correction and maximized $Z^2_1$ over the orbital phase and frequency derivative at the candidate spin frequency. The dashed white rectangle marks the inset region. The color scale gives the local Leahy $Z^2_1$ power; since the coherent $H$-test maximum occurs at $m=1$, this map presents the corresponding single-harmonic view of the candidate solution. The white contours mark approximate 1, 2, and 3$\sigma$ levels using $\Delta Z^2_1=2.30$, 6.18, and 11.83 for two profiled parameters.}
\label{fig:banana_htest}
\end{figure*}

{
Figure~\ref{fig:banana_htest} shows the corresponding single-harmonic Leahy/Rayleigh banana-plot in the $P_{\rm orb}$--$a_1\sin i$ plane. At each grid point, the photon arrival times were corrected with a circular-orbit delay model, and the plotted value is the maximum $Z^2_1$ power obtained after profiling over the orbital phase and frequency derivative at the candidate spin frequency. The map shows a ridge around the candidate solution rather than an isolated single-pixel maximum, illustrating the covariance between $P_{\rm orb}$ and $a_1\sin i$ over the limited observing baseline. Because the full $H$-test solution is maximized at $m=1$, this $Z^2_1$ map is the corresponding single-harmonic view of the coherent-search result.
}

{
We applied the same PSO-based coherent-search framework independently to the 2018 \textit{XMM-Newton}/PN data, using the $H$-test as the objective function. The explored parameter space in
$(a_1, P_{\rm orb}, T_{\rm asc}, \nu, \dot{\nu})$ was identical to that adopted for the 2023 observation, and no orbital parameters from the 2023 solution were used as priors in the 2018 search.
}

{
Within this parameter space, a candidate coherent solution is recovered at a frequency of $\nu \simeq 0.223$~Hz, where the $H$-test reaches a local maximum of $H_{\rm max}=47.2$. For this solution, the $H$-test is maximized at $m=3$, corresponding to a $Z^2_3$ value of 55.2.
}

{
In the 2018 observation, the available photon statistics are substantially lower ($N_{\gamma}=4378$ for 0.2--12.0~keV and $N_{\gamma}=3116$ for 0.5--12.0~keV), which likely contributes to the reduced detection strength. Given the lower photon counts and the limited observing baseline, we do not attempt to independently constrain the orbital parameters from the 2018 dataset alone, and we treat this result as a consistency check rather than as a standalone candidate detection.
}

{
As a confirmation of the recovered phase coherence, we computed the Leahy-normalized Fourier power spectra of the orbit-corrected time series. To illustrate how the pulsation power is distributed among harmonics, we constructed Fourier-domain representations equivalent to the $Z^2_n$ statistics with $n=1,2,3$, by coherently summing the Leahy power at the fundamental frequency and its harmonics on a uniform frequency grid.
}

\begin{figure*}[t]
\centering
{
\includegraphics[width=\textwidth]{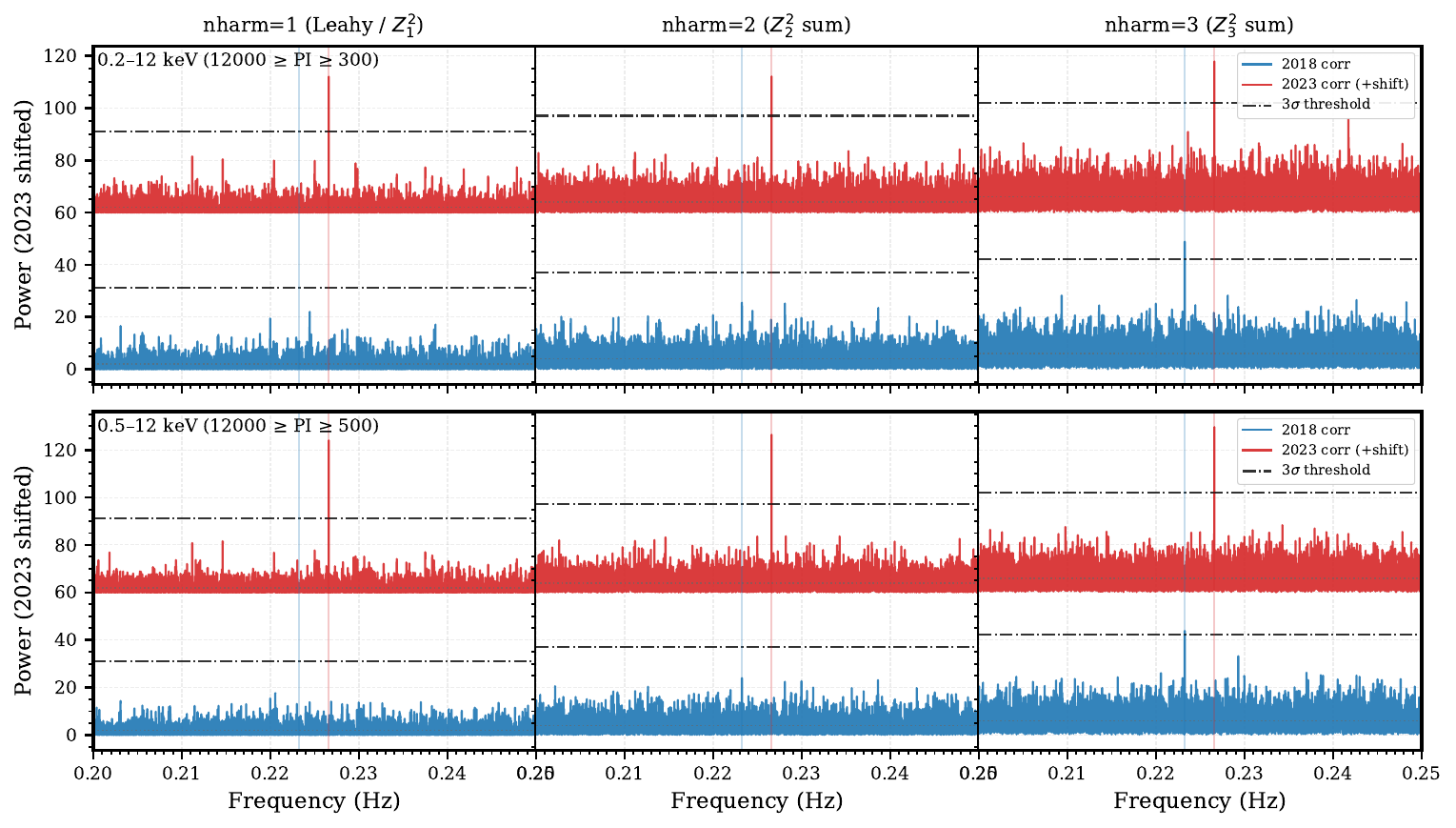}
}
\caption{{
Fourier-domain representations equivalent to the $Z^2_n$ statistics ($n=1,2,3$) computed from the Leahy-normalized power spectra of the orbit-corrected time series for the 2018 and 2023 \textit{XMM-Newton}/PN observations.
The top and bottom rows show the 0.2--12.0~keV and 0.5--12.0~keV selections, respectively. For clarity, the 2023 spectra are shifted upward by a constant offset of 60. A narrow and dominant peak at the pulsation frequency is recovered for all $n$, demonstrating that the orbital demodulation restores near-complete phase coherence across the observation.
}}
\label{fig:harmonics_fft}
\end{figure*}

{
Figure~\ref{fig:harmonics_fft} shows these $Z^2_n$-equivalent spectra for the 2018 and 2023 observations in two energy selections (0.2--12.0~keV and 0.5--12.0~keV).
For visual clarity, the 2023 curves are shifted upward by a constant offset of 60. A narrow and dominant peak is present at the pulsation frequency in 2023 spectra, demonstrating that the orbital demodulation successfully restores near-complete phase coherence across the observation.
}

{
We emphasize that the PSO algorithm is employed solely as an efficient numerical optimizer to locate the maximum of the $H$-test statistic in a high-dimensional parameter space.
Given the limited observing baseline ($T_{\rm obs} \simeq 124~\mathrm{ks}$), the inferred orbital parameters should be regarded as an effective phase-correction solution rather than a unique physical orbital determination. Nevertheless, the substantial and consistent coherence recovered after demodulation supports the interpretation that the candidate pulsation signal is intrinsic to NGC~7456~ULX--1.
}

\subsection{Pulse Profile and Energy Dependence}

{
Using the orbit-corrected event lists obtained in Section~3.2, we extracted pulse profiles for both the 2018 and 2023 \textit{XMM-Newton}/PN observations by folding the photon arrival times at the best-fit spin frequency. All profiles shown here are background-subtracted and normalized by the mean count rate.
}

\begin{figure*}[t]
\centering
{\includegraphics[width=0.55\textwidth]{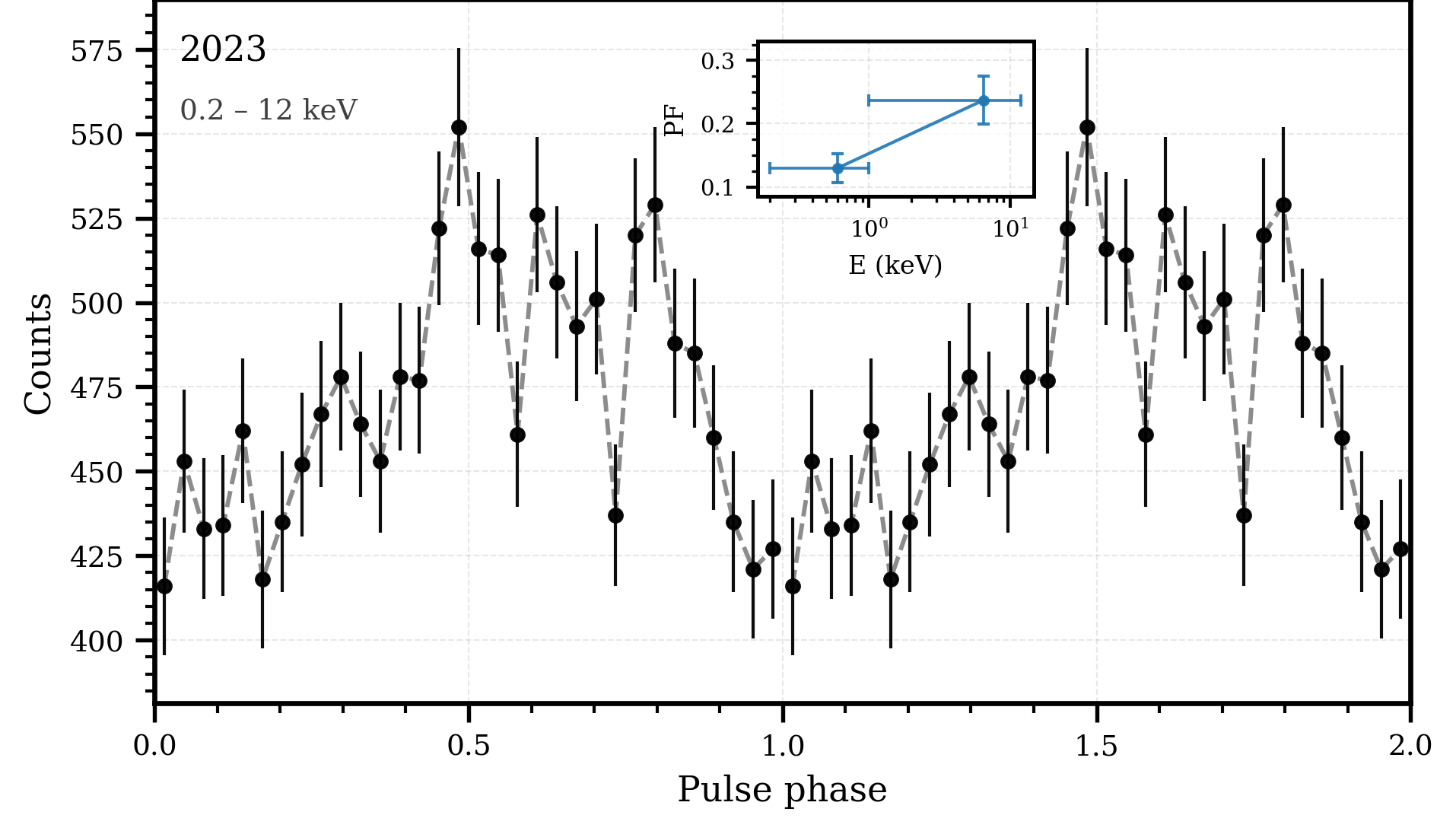}}
\caption{
Orbit-corrected pulse profile of NGC~7456~ULX--1 in the 0.2--12.0~keV band for the 2023 \textit{XMM-Newton}/PN observation. Two pulse cycles are shown for clarity. A broad candidate modulation is visible in the data. The inset displays the energy dependence of the pulsed fraction for the same observation.
}
\label{fig:pulse_profiles}
\end{figure*}

{
Figure~\ref{fig:pulse_profiles} displays the pulse profile in the 0.2--12.0~keV band in the 2023 observation\footnote{The lower-significance 2018 profile is not shown because it is strongly limited by the smaller photon statistics and is not treated as an independent pulsation detection.}.
It exhibits a clear broad modulation that is approximately sinusoidal. The folded profile obtained from the 2018 dataset is strongly limited by the lower photon statistics, and we therefore do not regard it as an independent pulsation detection, but only as a low-significance consistency check under the same demodulation procedure.
}

{
To investigate the energy dependence of the pulsation, we further extracted energy-resolved pulse profiles in several bands. In the 2023 observation, the pulse shape remains broadly sinusoidal at low energies and becomes more peaked toward higher energies, suggesting that the pulsed emission is more strongly associated with the hard X-ray component. In the 2018 dataset, however, the corresponding energy-resolved profiles are not sufficiently constraining to establish a comparable trend on their own.
}

{
The energy dependence of the pulsed fraction in the 2023 observation is shown in the inset of Figure~\ref{fig:pulse_profiles}. The pulsed fraction increases monotonically with energy, rising from $\sim0.1$--$0.15$ below 1~keV to $\gtrsim0.25$ above a few keV.
Such an increase of pulsed fraction with energy is a well-established property of ultraluminous X-ray pulsars and is commonly interpreted as evidence that the pulsed emission originates predominantly from the accretion column or magnetically channeled flow \citep[e.g.,][]{bachetti_ultraluminous_2014, walton_potential_2018}.
}

{
Overall, the 2023 pulse-profile properties are compatible with an accretion-powered neutron-star interpretation, whereas the 2018 observation remains too limited to provide an independent constraint.
}

\subsection{Global false-alarm probability}
\label{subsec:fap}

To quantify the statistical significance of the coherent-search result, we evaluate the global false-alarm probability under a noise-only null hypothesis,
\[
\mathrm{FAP}=P\!\left(S_{\max}\ge S_{\rm obs}\mid H_0\right),
\]
where $H_0$ denotes the adopted noise-only model, $S_{\rm obs}=64.6$ is the observed maximum detection statistic, and
\[
S_{\max}=\max_{\theta\in\Theta} H(\theta),
\]
with $\theta=(a_1,P_{\rm orb},T_{\rm asc},\nu,\dot{\nu})$ and $\Theta$ the full parameter domain explored in the coherent search. The relevant null distribution is that of the maximum statistic returned by the complete search, not the distribution of $H$ at a pre-specified point, because the reported value is selected after optimizing over the five-dimensional parameter space. This is the usual look-elsewhere problem for period searches and other multi-template searches; we therefore calibrate $S_{\max}$ directly with end-to-end simulations and use extreme-value modeling only to describe the unresolved far tail of the simulated maxima distribution \citep[e.g.,][]{baluev_assessing_2008,suveges_extreme_value_2014}.

We estimate the global false-alarm probability (FAP) with Monte Carlo simulations that reproduce the full orbit-corrected coherent-search procedure. Specifically, we generate 100,000 noise-only event lists by randomly redistributing the photon arrival times within the good time intervals, while preserving both the total number of events and the GTI structure. Each simulated event list is then analyzed with the same PSO-based coherent search over the full parameter space $\Theta$ as adopted for the real data, and we record the maximum $H$-test value returned by that search. The resulting simulated distribution therefore corresponds to the null distribution of the full-search maximum and directly accounts for the look-elsewhere effect within the adopted search setup. We do not rely on an explicit analytic trial-factor correction based on a naive template count, because in the present five-dimensional coherent search the orbital and spin parameters are strongly correlated over the limited observing baseline, so the nominal number of parameter-space templates would substantially overestimate the number of statistically independent trials. For computational efficiency, the PSO settings used in the Monte Carlo runs are moderately reduced relative to the production search, while keeping the same objective function and the same parameter bounds. None of the 100,000 noise realizations yields a maximum statistic exceeding the observed value ($S_{\rm obs}=64.6$), implying a conservative empirical upper bound of $\mathrm{FAP}_{\rm emp}<10^{-5}$.

\begin{figure}[t]
\centering
\includegraphics[width=0.8\columnwidth]{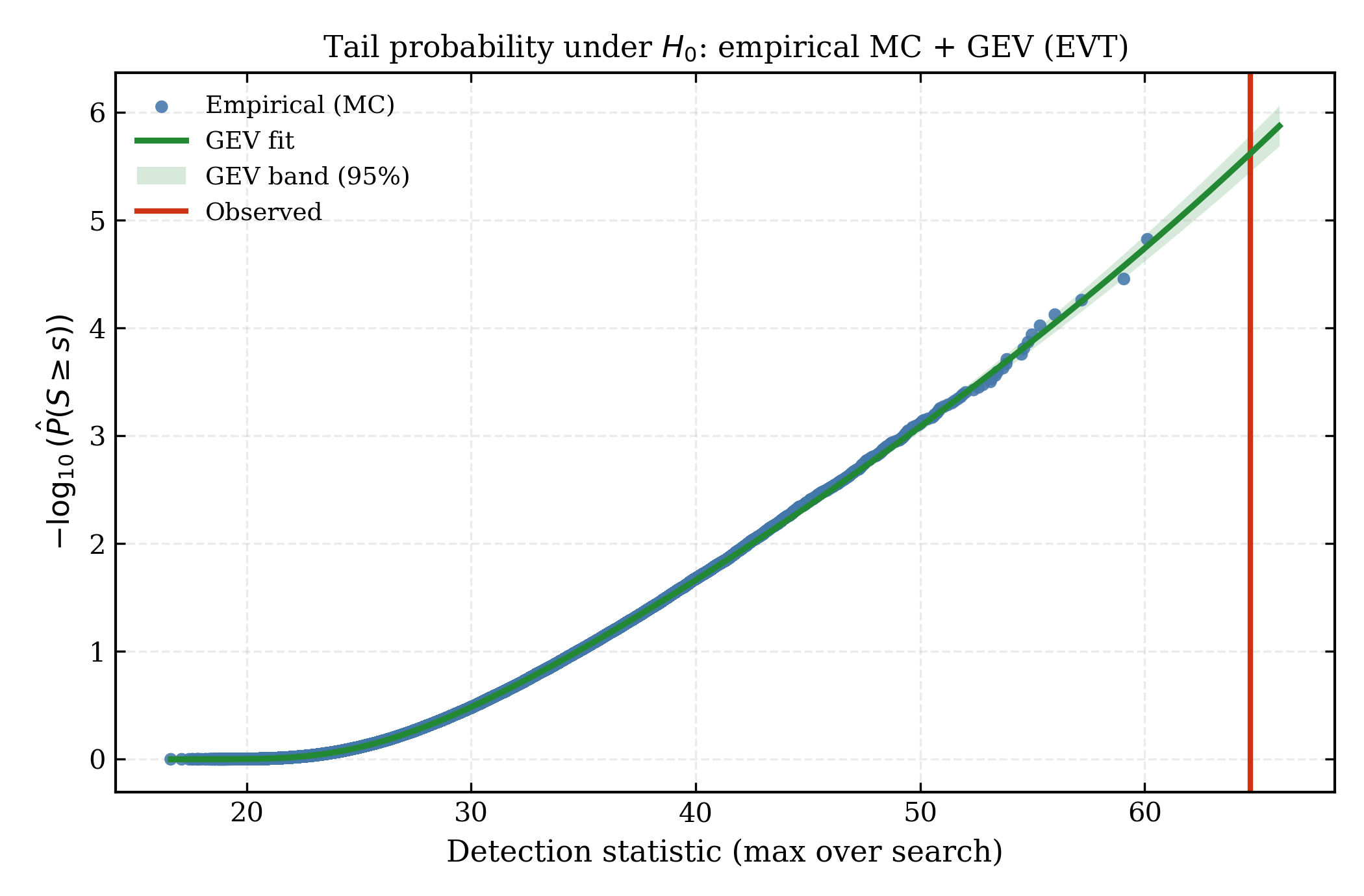}
\caption{{
Empirical distribution of the maximum detection statistic obtained from 100,000 noise-only Monte Carlo realizations of the full PSO-based coherent search applied to the 2023 data. The solid curve shows the best-fit generalized extreme value (GEV) model to the simulated maxima distribution. The vertical red line marks the observed value from the real data ($S_{\rm obs}=64.6$), which lies beyond the range directly sampled by the Monte Carlo simulations, with no simulated realization exceeding this value. Extrapolation of the fitted GEV model to the observed statistic gives an estimated tail false-alarm probability of $\mathrm{FAP}_{\rm GEV}\approx 2.5\times10^{-6}$.
}}
\label{fig:gev_tail}
\end{figure}

Because the observed value lies beyond the range sampled directly by the finite Monte Carlo set, we further use Extreme Value Theory \citep[EVT;][]{baluev_assessing_2008,suveges_extreme_value_2014} to characterize the far tail of the simulated maxima distribution. We fit the distribution of the simulated search maxima with a generalized extreme value (GEV) model and use it only as an auxiliary description of the extreme tail beyond the empirical Monte Carlo range. Figure~\ref{fig:gev_tail} shows the empirical distribution of the maximum detection statistic from 100,000 noise-only Monte Carlo realizations together with the best-fit GEV model. For the observed value, $S_{\rm obs}=64.6$, no simulated realization exceeded the observed statistic, so the empirical tail probability is not directly resolved by the Monte Carlo sample itself. Under the fitted GEV description of the tail, the observed statistic corresponds to an estimated global false-alarm probability of $\mathrm{FAP}_{\rm GEV}\approx 2.5\times10^{-6}$. As an additional empirical check, the analyses of ULX--2 on the same detector and of source-free background regions do not show a comparably significant signal, although the ULX--2 test is limited by its smaller photon statistics. These results argue against a noise, background, or detector-wide origin for the candidate signal.

However, the plausibility of the candidate signal does not imply that the orbital parameters associated with the optimal demodulation represent a unique physical solution. Because the observing span is comparable to at most a single orbital cycle, the recovered parameters are better regarded as the combination that maximizes phase coherence within the present dataset. Independent confirmation with additional observations or other instruments is still required to verify the signal and to better constrain the orbital parameters.

\section{Discussion and Conclusions}

We find evidence for a candidate coherent pulsation signal in NGC 7456 ULX-1. The candidate is primarily established through a fully coherent timing analysis of the 2023 \textit{XMM-Newton} observation, in which orbital demodulation restores phase coherence across the exposure and yields a pronounced peak in the detection statistic. Using both the \texttt{HENZSEARCH} algorithm in \texttt{HENDRICS} and the \texttt{ACCELSEARCH} routine within \texttt{PRESTO}, we identify a candidate periodic signal near 0.22 Hz with consistent local significance estimates. Although blind and acceleration searches are not optimal in the presence of orbital Doppler modulation and rapid spin evolution, the recovery of phase coherence after orbital demodulation, together with dedicated Monte Carlo tests, indicates that the observed peak is unlikely under the adopted noise-only hypothesis. These results suggest that NGC 7456 ULX-1 may host an accreting neutron star that produces a candidate pulsation signal, although independent confirmation with additional observations is still required.

The observed spin evolution of the candidate signal, if confirmed, would provide constraints on the surface magnetic field strength ($B$) of the putative neutron star.
Assuming that the spin evolution is driven by accretion torque, we apply the following relation \citep{Frank2002}
\begin{equation}
    2\pi I\dot{\nu}\simeq\dot{M}(GMR_{\rm in})^{1/2},
\end{equation}
where $I$, $M$, and $\dot{M}$ are the moment of inertia, mass, and mass accretion rate of the neutron star respectively, $G$ is the gravitational constant, and $R_{\rm in}$ is the inner radius of the accretion disk. Equation (1) can be rewritten in the following form
\begin{equation}
    \dot{\nu}\simeq 3.3\times 10^{-11}\,\eta^{1/2}I_{45}^{-1}m_1^{6/7}\dot{m}^{6/7}\mu_{30}^{2/7}\,{\rm Hz\,s}^{-1},
\end{equation}
where $\eta$ is the ratio between $R_{\rm in}$ and the Alf\'ven radius in spherical accretion, which is of order unity \citep{ghosh_accretion_1979}, $I_{45}=I/10^{45}\,{\rm g\,cm}^2$, $m_1=M/1\,M_\odot$, $\dot{m}=\dot{M}/\dot{M}_{\rm Edd}$, and $\mu_{30}=\mu/10^{30}\,{\rm G\,cm}^3$ is the normalized magnetic moment.

Analyzing the temporal evolution of the pulse period from 2018 to 2023, we find an average spin-up rate of $\dot{P}\sim -1.3 \times 10^{-8}$ s\,s$^{-1}$, equivalent to $\dot{\nu}\sim 10^{-9}$ Hz\,s$^{-1}$. The extremely high luminosities ($L_{\rm X} \sim 10^{40}$ erg\,s$^{-1}$) indicate super-Eddington accretion, implying that the inner disk likely lies within the spherization radius ($R_{\rm sph}$) as proposed by \citet{Shakura1973}. In this regime, excess mass flow within $R_{\rm sph}$ is expelled in the form of a wind. Under such circumstances,  the spin-up rate can be expressed as \citep{King2017}.
\begin{equation}
    \mu_{30}\simeq 1.3\eta^{-7/4}I_{45}^{3/2}m_1^{1/2}\dot{\nu}_{-9}^{3/2},
\end{equation}
where $\dot{\nu}_{-9}=\dot{\nu}/10^{-9}$ Hz\,s$^{-1}$.

On the other hand, the neutron star could be around spin equilibrium. From the value of the equilibrium period \citep{king_ultraluminous_2023}
\begin{equation}
    P_{\rm eq}\simeq 0.23\eta^{7/6}m_1^{-1/3}\mu_{30}^{2/3}\,{\rm s}\sim 5\,{\rm s},
\end{equation}
we estimate $\mu_{30}\sim 100$.

Thus, the timing properties inferred from the candidate signal are consistent with NGC 7456 ULX-1 hosting a strongly magnetized neutron star, with an estimated field strength of $B\sim 10^{12}$--$10^{14}$~G, undergoing rapid accretion.

\begin{acknowledgements}
We are grateful to an anonymous referee for useful comments that helped improve the manuscript. This work was supported by the National Key Research and Development Program of China (2021YFA0718500) and the Natural Science Foundation of
China under grant Nos. 12121003 and 123B2045.
\end{acknowledgements}

\bibliography{ref}{}
\bibliographystyle{aasjournal}

\end{document}